 \documentclass[final,5p,times,twocolumn,authoryear]{elsarticle}

\usepackage{amssymb}
\usepackage{lipsum}
\usepackage{changepage}
\usepackage{import}
\usepackage{amsmath}
\usepackage{tabularx,multirow,booktabs,blindtext}
\usepackage{adjustbox}
\usepackage{subfig}
\usepackage{placeins} 
\usepackage{hyperref}
\usepackage{graphicx,rotating}
\usepackage{mwe}
\usepackage{txfonts}
\usepackage[T1]{fontenc}
\usepackage{epsfig}
\usepackage{natbib}
\bibpunct{(}{)}{;}{a}{}{,} % to follow the A&A style
\usepackage{color}
\usepackage{amssymb}  
\usepackage{tabularx}
\usepackage{tabulary}
\usepackage{comment}
\usepackage{caption}
\usepackage{subcaption}
\usepackage{graphicx}
\usepackage{txfonts}
\journal{Astronomy $\&$ Computing}

\begin{document}

\begin{frontmatter}



\title{A Guided Unconditional Diffusion Model to Synthesize and Inpaint Radio Galaxies from FIRST, MGCLS and Radio Zoo}

%%             postcode={Bellville 7535},

\author[first]{R. Poitevineau}
\author[first]{E. Tolley}
\author[second]{V. Etsebeth}
\affiliation[first]{{Institute of Physics, Laboratory of Astrophysics, Ecole Polytechnique Federale de Lausanne (EPFL)},%Department and Organization
            addressline={1290 Sauverny}, 
            state={Switzerland}}
\affiliation[second]{{Department of Physics and Astronomy, University of the Western Cape},%Department and Organization
            addressline={Private Bag X17}, 
            postcode={Bellville 7535},
            state={South Africa}}

\begin{abstract}
We present a masked-guided approach for a denoising diffusion probabilistic model (DDPM) trained to generate and inpaint realistic radio galaxy images. The inpainting capability is particularly relevant for reconstructing incomplete observations, improving downstream tasks such as source characterization and morphological classification. We train the model on a combination of the FIRST survey, Radio Galaxy Zoo, and cutouts from the MGCLS survey, enabling it to capture a broad range of radio galaxy morphologies across different observational regimes.
We evaluate the realism of the generated samples through statistical comparisons with real data, ensuring consistency in key morphological and intensity distributions. Our unconditional model produces morphologically plausible galaxies while maintaining diversity, highlighting the suitability of diffusion models for this task. This approach provides a scalable alternative to computationally expensive simulations and enables effective data augmentation for machine learning applications in radio astronomy, including source detection, classification, and image reconstruction.
\end{abstract}




\end{frontmatter}

%\tableofcontents



\section{Introduction}
\label{introduction}

The Square Kilometer Array (SKA) is expected to significantly advance radio astronomy by providing an unprecedented volume of high-resolution observational data. With an estimated output of 600 PB of calibrated data products per year~\citep{skao}, SKA will enable deeper insights into radio-emitting astrophysical phenomena. In particular, it will contribute to the study of radio galaxies and galaxy clusters, including the impact of active galactic nucleus (AGN) feedback on galaxy evolution, as well as the properties of cosmic rays and magnetic fields in the intracluster medium (ICM).

Radio galaxies and galaxy clusters play a crucial role in shaping the evolution of cosmic structures. Radio-loud AGNs drive powerful jets that release vast amounts of energy, injecting heat and momentum into their environment. This process influences both the properties of their host galaxies and the surrounding ICM \citep[e.g.,][]{2013kormendy}. On larger scales, these jets contribute to AGN-driven feedback, which regulates the thermal and dynamical state of the ICM, particularly within galaxy clusters \citep[e.g.,][]{2008Miley_DeBreuck, 2012fabian, 2022Magliocchetti}.

Cosmic rays accelerated by AGN jets, along with shocks induced by galaxy mergers, generate diffuse radio emission in the form of radio halos and relics \citep{2019vanweeren,2021zuhone}. These non-thermal components of the ICM are strongly influenced by the presence of magnetic fields, which shape the trajectories of charged particles and play a fundamental role in energy transport within galaxy clusters. Investigating the interactions between AGN activity, cosmic rays, and magnetic fields provides essential insights into the mechanisms driving galaxy evolution and the formation of large-scale cosmic structures \citep{2023wittor}.

However, the scale and complexity of SKA datasets will render manual analysis impractical, necessitating the development of automated detection and classification methods. Machine learning (ML) techniques provide a promising solution due to their ability to efficiently process large datasets and model complex relationships within data. Among these, generative models offer a valuable approach by producing synthetic data that can serve as empirically driven mock observations, which can be used to augment training datasets, improve model generalization, and facilitate robust uncertainty quantification~\citep{2023ribas,2024doorenbos}.

Generative models have proven to be powerful tools in various image-processing applications, including super-resolution, inpainting, and data augmentation. In super-resolution, models such as generative adversarial networks (GANs) and diffusion models enhance low-resolution images by generating high-resolution details, improving clarity and sharpness~\citep{2016ledig,2023feng,2025bachimanchi}. Inpainting techniques use generative models to restore missing or corrupted parts of an image, seamlessly reconstructing gaps by learning contextual structures~\citep{2022andreas,2023wang}. In astrophysics, generative models have been applied to parameter inference for stellar populations~\citep{Laroche_2025}, simulating galaxy observations~\citep{2016aravangakhsh,2021spindler}, and analyzing microlensed gravitational waves~\citep{2024bada}. GANs, in particular, have been used for denoising astrophysical data, enhancing gravitational wave detection~\citep{2023Murali}, deconvolution~\citep{2017schawinski}, synthetic galaxy image generation~\citep{2019fussel}, and radio halo detection~\citep{2024ashutosh}.

Despite the success of ML in various astronomical applications, its use in radio astronomy, particularly for radio galaxies, presents distinct challenges. Radio galaxies exhibit complex and diverse morphologies, often characterized by diffuse emission structures and extended lobes. Unlike optical galaxies, which typically have well-defined features~\citep{2022smith}, radio galaxies are affected by noise and instrumental artifacts, complicating automated analysis~\citep{2024URSLTolley}. Furthermore, the limited availability of annotated datasets in radio astronomy hinders the development of robust ML models. Addressing these challenges requires generative models capable of producing realistic synthetic datasets to supplement existing labeled data.

Among generative models, the most common approaches include variational autoencoders (VAEs)\citep{2013kingma}, GANs\citep{2014goodfellow}, and denoising diffusion probabilistic models (DDPMs)\citep{2015sohl-dickstein}. While GANs produce sharp images efficiently, they suffer from training instabilities and mode collapse. VAEs offer stable training and efficient inference but often generate blurry images due to their variational loss. DDPMs, by contrast, effectively capture a wide range of data distributions while avoiding these issues, enabling the reconstruction of fine details with high fidelity\citep{2020ddpm}.

DDPMs generate high-quality, diverse images by iteratively refining noise, ensuring strong mode coverage and stability. This makes them particularly well-suited for simulating a broad range of radio galaxy morphologies, including extended lobes and diffuse emission, thereby enriching the diversity and utility of synthetic datasets. Such improvements in data generation can enhance the training and performance of ML models for radio astronomy applications. However, DDPMs are computationally expensive due to their iterative denoising process, requiring multiple forward passes per image, which makes them significantly slower than GANs.

Each generative model presents trade-offs between quality, stability, and computational efficiency, making it crucial to evaluate their suitability for specific astrophysical applications. Given the importance of preserving fine structural details in radio galaxy images, DDPMs stand out as a promising approach despite their higher computational cost.

DDPMs have recently been applied in astrophysics, with \citet{2022smith} training a DDPM on data from the Dark Energy Spectroscopic Instrument \citep{2019AJ....157..168D} to generate synthetic observations tailored to a specific survey. Similarly, \citet{2024Vicanek} developed conditional models trained on cutouts from the LOFAR Two-Meter Sky Survey \citep{2017LOFAR} and on the Faint Images of the Radio Sky at Twenty-Centimeters survey (FIRST) \citep{first2023}.

In this work, we focus on a masked guided diffusion process to help the network generalize information and be able to generate and inpaint realistic radio galaxies images without the need to retrain the model or perform multiple sampling per image inpainting \citep{2022inpaint}. Furthermore, we propose an unconditional approach using multiple survey catalogs to train a single model. In Section~\ref{sect:data}, we describe the dataset used to train our model. Section~\ref{sect:ddpm} introduces the DDPM formalism, followed by Section~\ref{sect:res}, where we detail the model architecture, ML framework, and data preprocessing, and present our results. Finally, we discuss our findings and future perspectives in Section~\ref{sect:conclu}.

%#######################################################
\section{Data}\label{sect:data}
%#####################
\subsection{FIRST}

The FIRST radio galaxies catalog is a consolidated dataset that integrates information from several prior catalogs \citep{2017mira,2010gendre,2017capettia,2017capttib,2018baldi,2011proctor}, each of which characterizes radio galaxy sources observed in the VLA FIRST  \citep{1995firstcat}. The catalog comprises 2,158 images, each with dimensions of 300 × 300 pixels. To mitigate the impact of noise, all pixel values below three times the local root-mean-square (RMS) of the noise are clipped to this threshold.

The radio galaxies in this dataset are categorized into four morphological classes: Fanaroff-Riley I (FR I) and Fanaroff-Riley II (FR II) sources \citep{1974fanaroff}, compact sources, and Bent sources. FR I and FR II classifications are based on the distribution of radio emission along the jets. The compact class includes unresolved point sources, while the Bent class comprises sources where the angle between the jets deviates significantly from 180 degrees.

%#####################
\subsection{MGCLS}

The MeerKAT Galaxy Cluster Legacy Survey (MGCLS) \citep{2022Knowles} is a comprehensive project that uses the sensitivity and resolution of the MeerKAT telescope to study galaxy clusters. It includes observations of 115 galaxy clusters from the MeerKAT continuum legacy survey.

MGCLS acquired data in the L band, with a frequency range spanning 900--1670 MHz. In the survey, two main imaging products were produced, tailored for different scientific objectives: high resolution images with an angular resolution of approximately 7''–8'' and convolved images that have been smoothed to a lower resolution of 15''.

MGCLS sources were extracted using Python Blob Detection and Source Finder (\textsc{PyBDSF}) \citep{2015Mohan}, using the default threshold parameters. Using this configuration, a total of 62,587 sources were extracted from the high resolution images. To focus on resolved and diffuse sources only, a size-based filtering criterion was applied (Estebeth in prep). Specifically, sources with an area smaller than eight times the beam solid angle were excluded. This cut effectively removed compact sources, retaining those that are significantly extended relative to the instrumental beam. As a result, the dataset was reduced to 7,051 sources.

%#####################
\subsection{Radio Galaxy Zoo}

Radio Galaxy Zoo is a citizen science project that is part of the Zooniverse \footnote{https://www.zooniverse.org/} platform. It aims to help astronomers identify and classify radio galaxies by enlisting the help of the public to visually inspect and match radio emission with their corresponding host galaxies seen in infrared images.
The subset of Radio Galaxy Zoo data used here is based on the first public data release, known as Radio Galaxy Zoo Data Release 1 \cite{2025rgz}. This subset comprises 20,000 unlabeled radio galaxy images extracted from the Radio Galaxy Zoo Data Release 1 (RGZ) catalog, which contains over 98,000 sources primarily derived from the FIRST survey. Each image measures 150 by 150 pixels, corresponding to a 4.5' by 4.5' region of the sky centered on the associated radio source.

%#######################################################
\section{Denoising Diffusion Probabilistic Models}\label{sect:ddpm}

DDPMs are a class of generative models that synthesize data through a two-step process: forward diffusion and reverse generation. During the forward diffusion phase, a Markov chain progressively adds Gaussian noise to the data, systematically degrading its structure. In the reverse generation phase, the model is trained to iteratively remove the added noise, and learns to reconstruct coherent and realistic data samples from an initial state of random noise.

%#####################
\subsection{Forward Process}

\begin{figure}[h]
    \centering
     \includegraphics[width=0.9\linewidth,keepaspectratio]{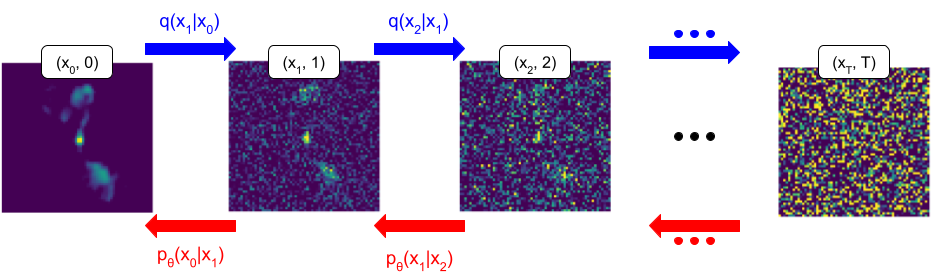}
    \caption{Illustration of the forward and backward process in a DDPM. The forward process (blue arrows) progressively adds Gaussian noise to an initial radio galaxy image $\mathrm{x_0}$, transforming it into a pure noise distribution $\mathrm{x_T}$ over $\mathrm{T}$ time steps. The backward process (red arrows) aims to learn the denoising operation $\mathrm{p_\theta(x_t \mid x_{t-1})}$, gradually removing noise to reconstruct the original image.} 
    \label{fig:noise_chain}
\end{figure}

Given a sample of images $x$, we can generate an individual image $x_0$ by sampling from a distribution $q(x)$. 
The forward process is defined as a sequential procedure where a small amount of Gaussian noise, $\mathcal{N}$, is gradually added to the input data over a series of steps, following a Markov chain formulation. 
This process generates a sequence of increasingly noisy representations, $\{x_1, x_2, ..., x_T\}$, where $T$ denotes the total number of steps in the forward process. Each step $x_t$ depending only on the previous state $x_{t-1}$. The posterior distribution for this process can be expressed as:
\begin{align}
q(x_{1:T} \mid x_0) &\coloneqq \prod_{t=1}^T q(x_t \mid x_{t-1}) \\
q(x_t \mid x_{t-1}) &\coloneqq \mathcal{N}(x_t ; \sqrt{1-\beta_t} x_{t-1}, \beta_t I)
\end{align}
where $\beta_t$ is a parameter that controls the variance of the noise added at each step, which can be learned by reparameterization \citep{2013kingma} or held constant as hyperparameters. A commonly used scheduler is based on a cosine function, which effectively controls the rate at which noise is added across the $T$ total steps \citep{2021alex}. 
The expressiveness of the reverse process is partly ensured by modeling the conditional $p_\theta(x_{(t-1)} \mid x_t)$ as a Gaussian distribution. This is because, when the noise schedule $\beta_t$ is small, the forward and reverse processes behave similarly, making the reverse process easier to learn \citep{2015sohl-dickstein}.

If we define the equation above to depend only on $x_0$, we can calculate any step $x_t$ explicitly \citep{2020ddpm}. Let us define $\alpha_t = 1 - \beta_t$ and $\overline{\alpha_t} = \prod^{t}_{i=1} \alpha_i$. Then, we can calculate:
\begin{align}
    x_t &= \sqrt{\alpha_t} x_{t-1} + \sqrt{1-\alpha_t} \mathcal{N}(0,I)\\
    &= \sqrt{\overline{\alpha_t}} x_0 + \sqrt{1 - \overline{\alpha_t}} \mathcal{N}(0,I)
\end{align}
Thus, we can also define $q(x_t)$ without any dependency on $x_{t-1}$:
\begin{equation}
    q(x_t \mid x_0) = \mathcal{N}(x_{t-1}; \mu_\theta(x_t,t), \Sigma_\theta(x_t,t))
\end{equation}

%#####################
\subsection{Backward Process}

Once the forward process is established, the objective of the diffusion model is to learn the reverse process, which progressively removes noise in a step-by-step manner to reconstruct the original data. This reverse process is also formulated as a Markov chain and is parameterized by a neural network.

The reverse process seeks to approximate $p_\theta (x_{t-1} \mid x_t)$, the conditional distribution of $x_{t-1}$ given $x_t$ where $\theta$ represents the parameters of the neural network. To achieve this, the model learns how to remove the noise contribution at each time step $t$, iteratively refining the data to recover a less noisy version $x_{t-1}$. This reverse transition is initialized by modeling $p_\theta (x_{t-1} \mid x_t)$ as a Gaussian distribution, given by:
\begin{equation}
p_\theta(x_{t-1} \mid x_t) = \mathcal{N}(x_{t-1}; \mu_\theta(x_t, t), \Sigma_\theta(x_t, t))
\end{equation}
where $\mu_\theta(x_t,t)$ is the mean of the conditional distribution, and $\Sigma_\theta(x_t, t)$ is the variance. The neural network takes the noisy input $x_t$ at time step $t$ and predicts $\mu_\theta(x_t,t)$, which guides the denoising process. The variance $\Sigma_\theta(x_t, t)$ represents the uncertainty in the reverse process and is often simplified or fixed during training to reduce computational complexity.
To calculate $x_{t-1}$, the model samples from the predicted Gaussian distribution $p_\theta (x_{t-1} \mid x_t)$: 
\begin{equation}
    x_{t-1} = \mu_\theta (x_t,t) + \sigma_t \epsilon
\end{equation}
where $\mu_\theta (x_t,t)$ is the predicted mean, $\sigma_t$ is the standard deviation derived from  $\Sigma_\theta(x_t, t)$, and $\epsilon$ is a sample from a standard normal distribution $\mathcal{N}(0,I)$. This step iteratively refines the data by removing the noise and recovering a less noisy version of the original input.

This parameterization enables the network to iteratively refine the data, leveraging the learned dynamics of the reverse process to accurately reconstruct the original distribution.

%#####################
\subsection{Architecture and training}\label{sect:archi}

\begin{table}[h]
\caption{\label{table:archi}Architecture}
\begin{center}
        \begin{tabulary}{\linewidth}{cccc}
                \toprule
                 Layer & Filters & Size & Stride\\
                \midrule  
                 Conv           & 128 & 3x3 & 1 \\
                 4 $\times$ Res & 128 & -   & - \\
                 Res1           & 128 & -   & - \\
                 Group Norm     &  8  & -   & - \\
                 Conv           & 256 & 2x2 & 2 \\
                 \cmidrule{1-4}

                 Group Norm     &  8  & -   & - \\
                 Conv           & 256 & 3x3 & 1 \\
                 4 $\times$ Res & 256 & -   & - \\
                 Res2           & 256 & -   & - \\
                 Group Norm     &  8  & -   & - \\
                 Conv           & 384 & 2x2 & 2 \\
                 \cmidrule{1-4}

                 Group Norm     &  8  & -   & - \\
                 Conv           & 384 & 3x3 & 1 \\
                 4 $\times$ Res & 384 & -   & - \\
                 Res3           & 384 & -   & - \\
                 Group Norm     &  8  & -   & - \\
                 Conv           & 384 & 2x2 & 2 \\
                 \cmidrule{1-4}

                 Group Norm     &  8  & -   & - \\
                 Conv           & 512 & 3x3 & 1 \\
                 5 $\times$ Res & 512 & -   & - \\
                 Group Norm     &  8  & -   & - \\
                 Conv           & 384 & 2x2 & 1 \\
                 \cmidrule{1-4}

                 Merge(Res3)    & -   & -   & - \\
                 Group Norm     &  8  & -   & - \\
                 Conv           & 384 & 3x3 & 1 \\
                 5 $\times$ Res & 384 & -   & - \\
                 Group Norm     &  8  & -   & - \\
                 Conv           & 256 & 2x2 & 1 \\
                \cmidrule{1-4}

                 Merge(Res2)    & -   & -   & - \\
                 Group Norm     &  8  & -   & - \\
                 Conv           & 256 & 3x3 & 1 \\
                 5 $\times$ Res & 256 & -   & - \\
                 Group Norm     &  8  & -   & - \\
                 Conv           & 128 & 2x2 & 1 \\
                \cmidrule{1-4}

                 Merge(Res1)    & -   & -   & - \\
                 Group Norm     &  8  & -   & - \\
                 Conv           & 128 & 3x3 & 1 \\
                 5 $\times$ Res & 128 & -   & - \\
                 Group Norm     &  8  & -   & - \\
                 Conv           & 128 & 1x1 & 1 \\

                \bottomrule
        \end{tabulary}
\end{center}
\caption{Architecture used of our DDPM. From left to right, we describe the layers name, the number of filters, the kernel size and the stride. The \textit{Conv} indicate convolutional layers, the \textit{Merge(Res)} indicate a layer where we concatenate the result of the \textit{Res} to the output of the previous layer. For the \textit{Group Norm} layers, the number of filters indicates the size of the groupe. The details of the Res layer is display in Table\ref{table:res_block}}.
\end{table}

The implementation was carried out inside the custom high performance deep learning framework CIANNA \citep{david_cornu_cianna}. The architecture adopts a U-shaped structure \citep{2015ronneberger}, a design widely used in encoder-decoder models for image processing tasks. It consists of sequential convolutional layers organized into downsampling and upsampling blocks. It takes images as inputs and produces images of same dimensions. In the downsampling blocks, the spatial dimensions of the input are reduced while the number of filters increases, enabling the extraction of hierarchical features. The upsampling blocks reverse this process, reconstructing the spatial dimensions while preserving learned features. Skip connections between corresponding downsampling and upsampling blocks are employed to retain fine-grained spatial information and enhance feature integration.

The architecture utilized in this study is detailed in Table~\ref{table:archi}. Each convolutional layer employs a leaky rectified linear unit with a slope of 0.1 for the activation function, except for the final layer, which uses a linear activation function.

In our architecture, max-pooling and unpooling operations were deliberately avoided for downsampling and upsampling, respectively. The primary rationale for this design choice lies in the need for consistency and adaptability in feature representation. Traditional unpooling operations typically require positional indices from the pooling layers in the encoder to restore spatial information, which introduces dependencies that may not generalize well for certain data types. Instead, transposed convolutions are used for upsampling, paired with convolutional operations for downscaling. This approach ensures a more homogeneous and expressive transformation across the network.

Moreover, the choice of transposed convolutions is suited for astronomical data, which generally exhibit smoother features and lack the sharp edges commonly found in natural images. This smoothness benefits from the learned upscaling process provided by transposed convolutions, which is expected to yield better performance in reconstructing fine details in astronomical imagery \citep{2024cornu}.

\subsection{Guided diffusion}\label{sec:guided}

\begin{figure}[h]
    \centering
     \includegraphics[width=0.9\linewidth,keepaspectratio]{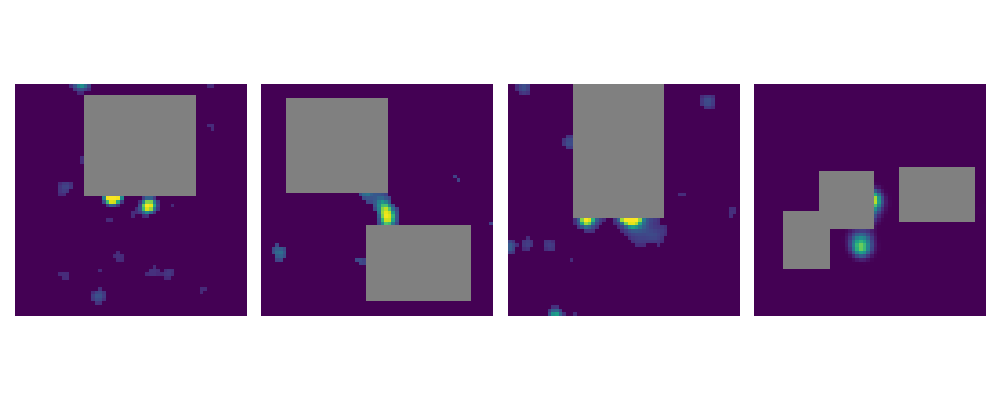}
    \caption{Example of some masked images used for training. The masked regions are represented in gray}
    \label{fig:masked_train}
\end{figure}

The model is trained for the inpainting task using images with randomly masked regions. Each training instance contains either a single rectangular mask occupying 40–80\% of the image area, two masks within the same size range, or three to four smaller masks occupying 20–40\% each (i.e, Figure~\ref{fig:masked_train}). To jointly enable unconditional image generation, a subset of training images is entirely masked. The proportion of fully masked samples is initially set to 20\% and is linearly increased to 80\% throughout the training process, thereby progressively shifting the model’s objective from conditional completion to full image synthesis.

By mixing partially masked and fully masked samples, the model learns both inpainting and unconditional generation diffusion processes within the same model.
Since the model must learn to infer both small- and large-scale missing content, it becomes more capable of leveraging global image context and semantic priors.
The benefit over a classic unconditional DDPM training regime lies in improved global structural coherence, enhanced semantic representation capacity, and greater diversity of conditional contexts during learning, resulting in sharper, more plausible full-image generations when the model is later sampled during the generation generation task \citep{2021glide,2023hdpainter}.

%#####################
\subsection{Data preprocessing}\label{sect:prepro}

To extract patches around sources in the FIRST dataset, we first identify regions of interest by applying an intensity threshold of 100 (on a 0–255 grayscale range), producing a binary mask that isolates significant emission from the background noise. On this mask, we perform connected-component labeling to detect contiguous pixel groups (“islands”), each corresponding to a candidate source or emission structure. For every detected island, we retrieve its full set of pixel coordinates and compute its bounding box, from which we estimate the geometric center. Fixed-size patches of 64×64 pixels are then extracted by centering on this position, while enforcing boundary conditions to ensure the patch remains fully contained within the image.
This initial extraction may produce multiple, potentially overlapping patches, especially in crowded regions. To address this, we rank all candidate patches by their total pixel intensity, used here as a proxy for source significance. We then apply a non-maximum suppression procedure based on the Intersection over Union (IoU): starting from the highest-intensity patch, we iteratively discard any remaining patches whose overlap exceeds a threshold of 0.2.
This process yields a final dataset consisting of 2,535 images.
We applied the same procedure to the RGZ data and obtained a sample of 20,468 images.

For the MGCLS cutouts, the images were first cropped to the nearest smaller power-of-two dimension relative to their original size. To ensure uniformity with the sample obtained from the FIRST dataset, larger images were subsequently resized to 64×64 pixels. Furthermore,
the background noise and flux levels were then calculated for each image and the pixel values below 3$\sigma$ were set to zero. The threshold of 3$\sigma$ was chosen to mitigate artifacts present in many cutouts due to insufficient calibration. Thus, we create data that should be more homogeneous with the FIRST data. The final MGCLS catalog contains 6915 images.

\begin{figure}[h]
    \centering
     \includegraphics[width=0.9\linewidth,keepaspectratio]{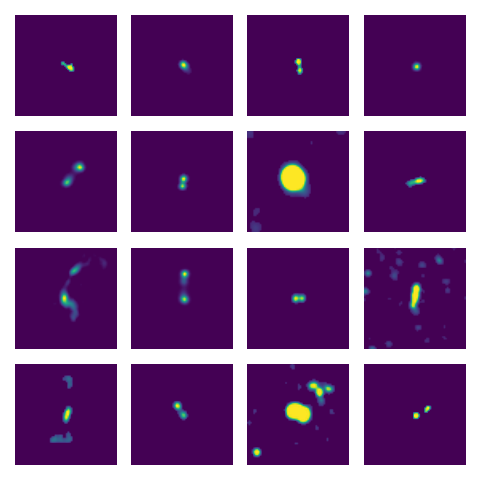}
    \caption{Example of real images used to train our trained DDPM.}
    \label{fig:ex_data}
\end{figure}

Combining our three datasets from FIRST, MGCLS and RGZ, we have a final sample of 29,918 radio galaxies. The final sample was normalized to a range between 0 and 1. To expand the dataset, data augmentation was applied, including rotations of 0, 90, 180, and 270 degrees, as well as horizontal and vertical flips. We show a few examples of these images in Figure~\ref{fig:ex_data}.

%#######################################################
\section{Results}\label{sect:res}

\begin{figure}[h]
    \centering
     \includegraphics[width=0.9\linewidth,keepaspectratio]{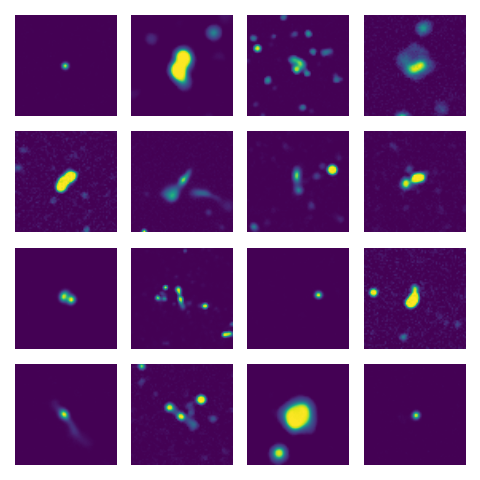}
    \caption{Example of images generated by our DDPM.}
    \label{fig:ex_gens}
\end{figure}

The DDPM was trained using the NVIDIA H100 GPU over 10000 epochs, employing the Adam optimizer \citep{2014adam}. Training the model required 24 hours, while generating 10,000 samples took 20 hours. The model has 67.7 miilions of parameters. We used a batch size of 32 and a learning rate of 0.0001. The sample generation was done without any semantic context, fully unconditional.

Metrics such as the Fréchet Inception Distance  and Inception Score are commonly used to evaluate the performance of DDPMs. However, these metrics rely on neural networks trained on everyday-life images, which differ significantly from astrophysical images, as discussed in Sect.~\ref{sect:archi}. Thus, we cannot be sure that the filters used in the neural networks are relevant to astrophysical images. Consequently, their applicability to assessing the performance of a DDPM trained on radio galaxy images is limited.

A more straightforward approach to evaluation is a qualitative visual comparison between the generated images and the real images used during training. We illustrate some examples of the generated images in Figure~\ref{fig:ex_gens}. As we can see, the generated radio galaxies look realistic compared to the real ones.

Beyond generating images that visually resemble radio galaxies, it is also essential that the synthesized images capture the underlying physical properties of the original data. These properties are reflected in the statistical distribution of pixel values. To assess the quality of the DDPM in this regard, we compare the statistical distributions of real and generated images. 

After training, the DDPM was used to generate 10,000 synthetic images, providing a sufficiently large sample for subsequent statistical analysis.

%#####################
\subsection{General statistics}

\begin{figure}[h]
    \centering
     \includegraphics[width=0.9\linewidth,keepaspectratio]{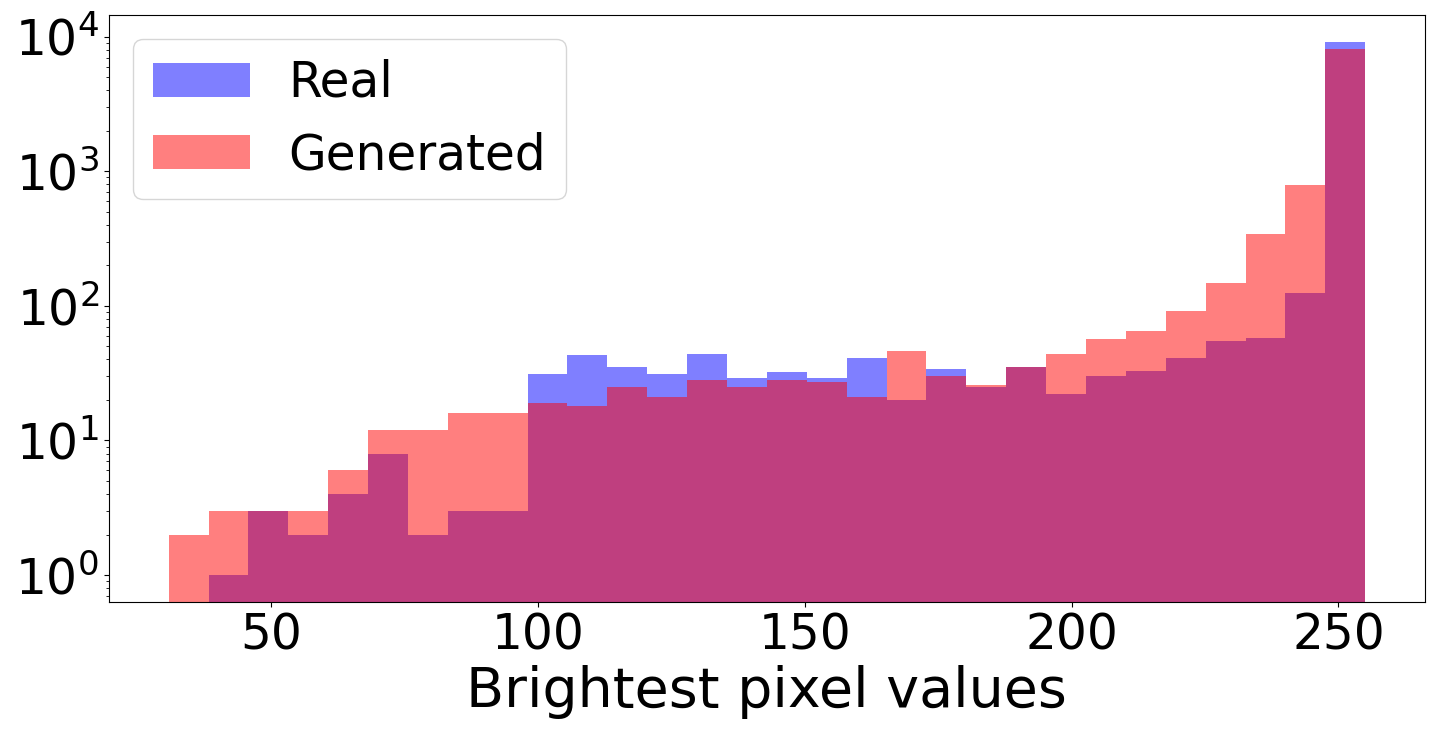}
    \caption{Statistical comparison of the brightest pixel values in each image in the real data sample and the generated one by our network}
    \label{fig:brightest_pix_val}
\end{figure}

\begin{figure}[h]
    \centering
     \includegraphics[width=0.9\linewidth,keepaspectratio]{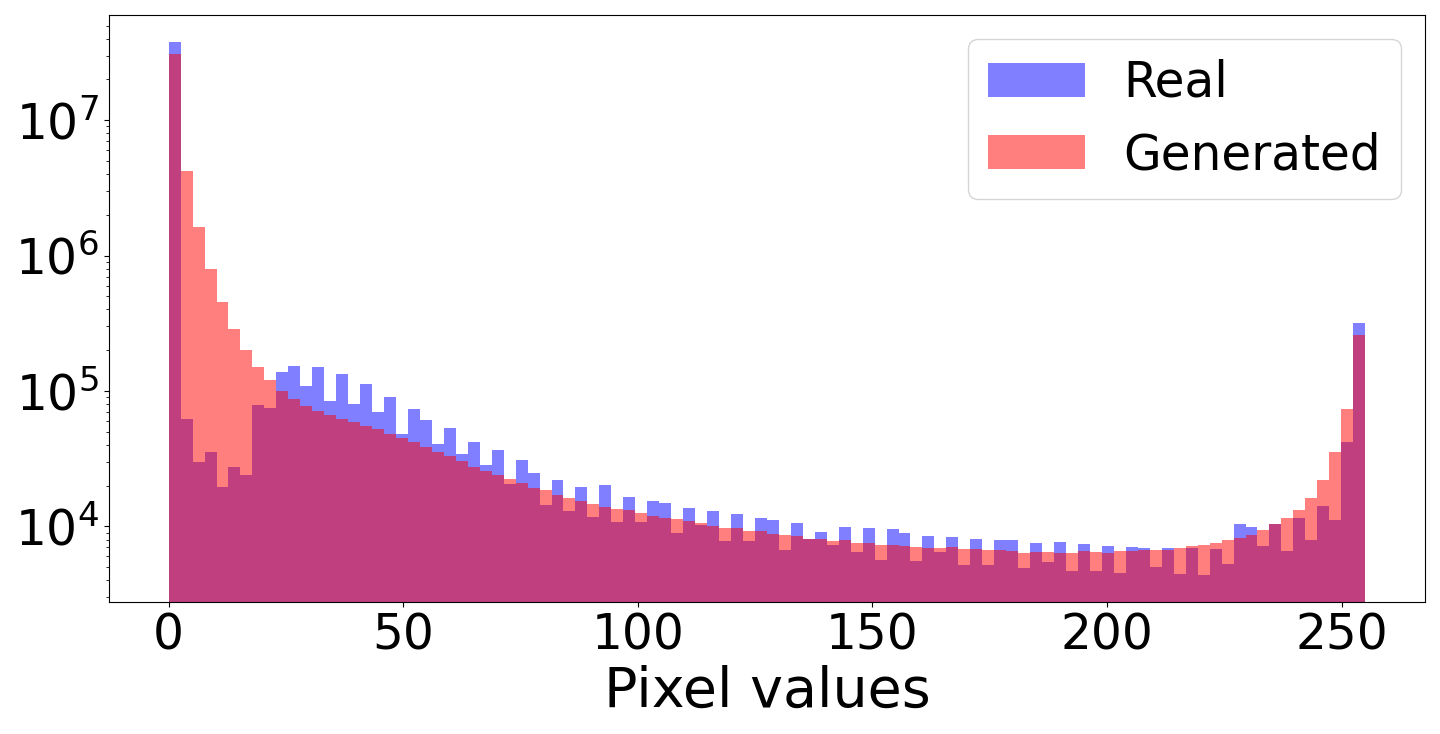}
    \caption{Statistical comparison of the pixel values distribution in the real data sample and the one generated by our network.}
    \label{fig:pix_val}
\end{figure}

\begin{figure}[h]
    \centering
    \includegraphics[width=0.9\linewidth,keepaspectratio]{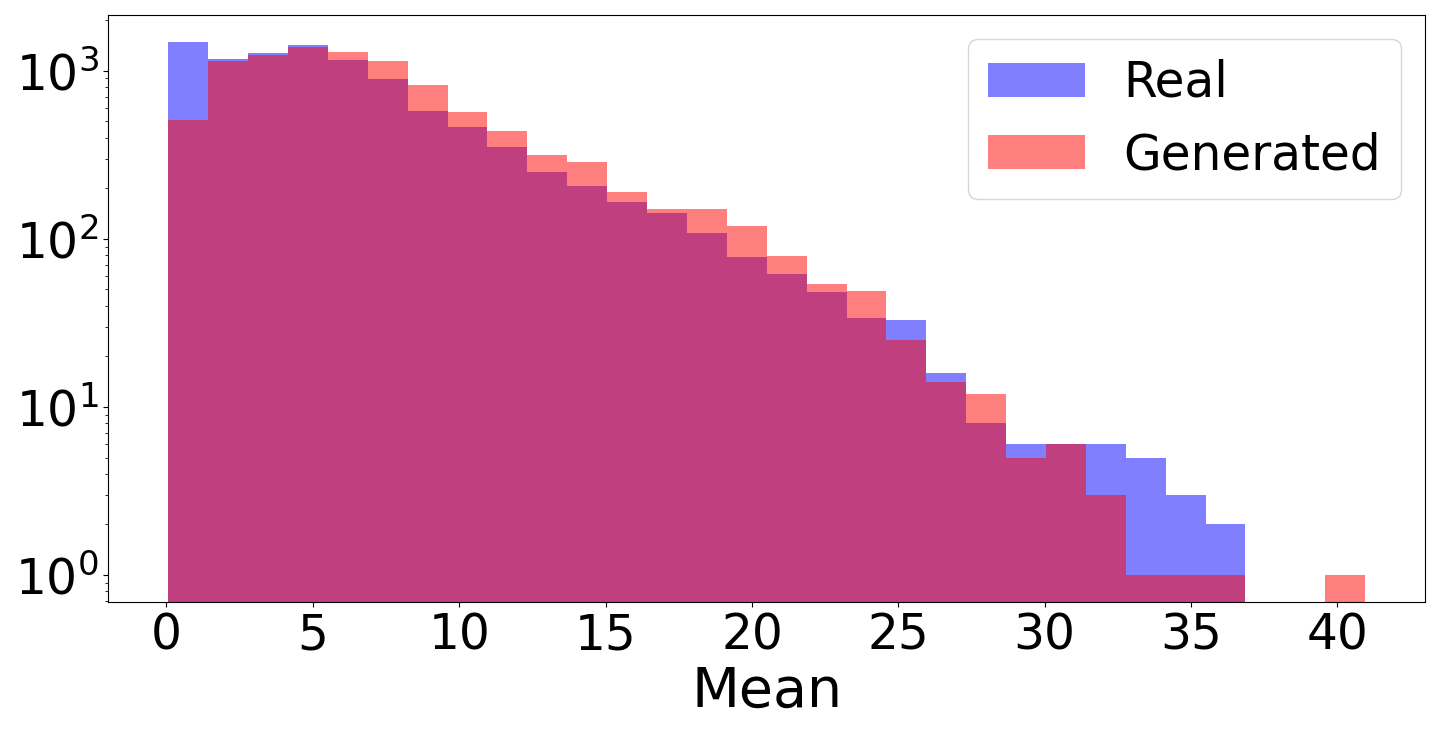}
    \caption{Statistical comparison of the mean pixel values in each image in the real data sample and the generated one by our network.}
    \label{fig:mean}
\end{figure}

\begin{figure}[h]
    \centering
    \includegraphics[width=0.9\linewidth,keepaspectratio]{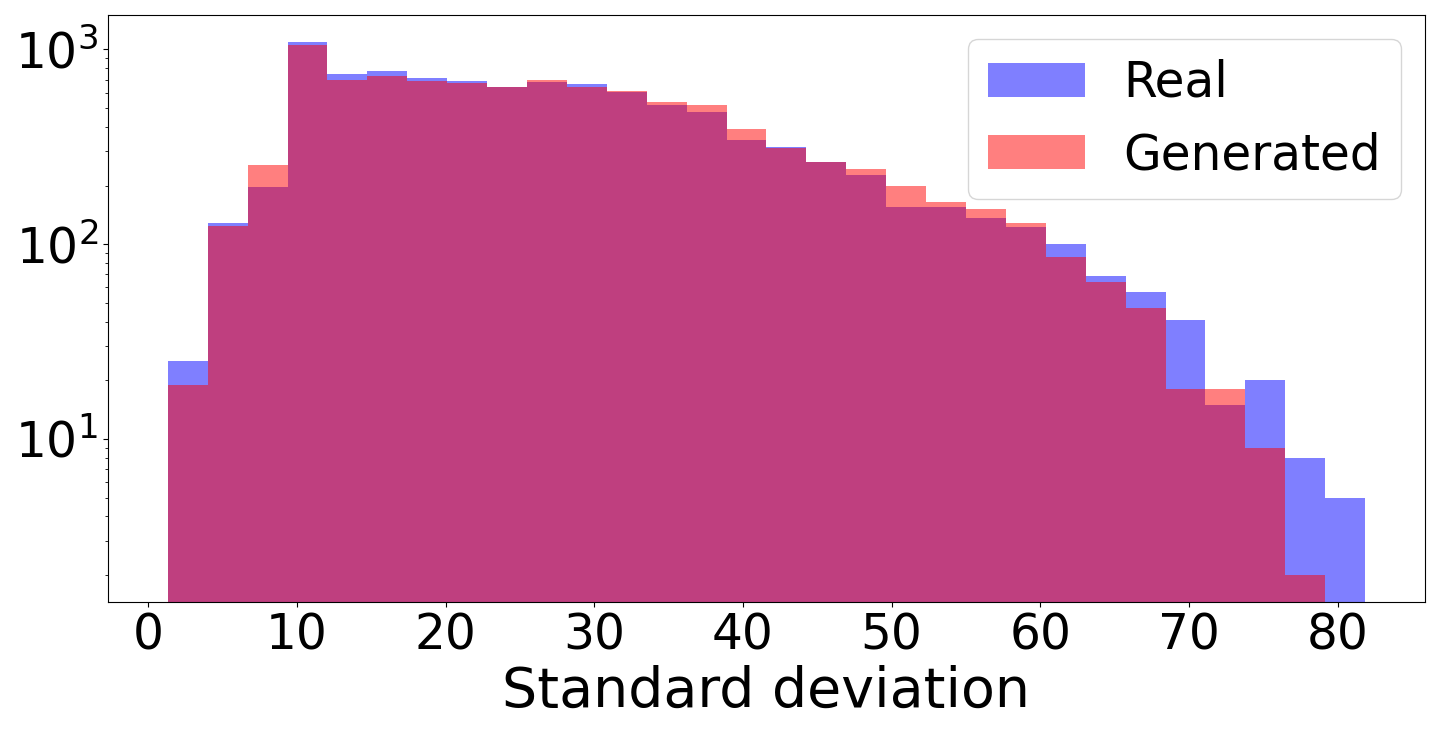}
    \caption{Statistical comparison of the standard deviation of the pixel values in each image in the real data sample and the one generated by our network.}
    \label{fig:std}
\end{figure}

To assess the performance of our network, we compare the statistical distributions of 10,000 generated sampled images with those of the 10,000 real data chosen randomly. 

Figure~\ref{fig:brightest_pix_val} presents the distribution of the brightest pixel values extracted from each image, comparing the real dataset (blue) with the generated samples (red). Both distributions extend from roughly 40 to the saturation limit near 255, and both display a strong increase in frequency toward the upper end of the dynamic range. The histograms overlap across most intensity bins, indicating that the model captures the general behavior of the brightest-pixel distribution. A noticeable difference appears at the highest intensity values, where the generated samples show a higher occurrence of saturated or near-saturated pixels. This suggests that the model slightly overestimates the frequency of very bright peaks compared to the real data, while still reproducing the overall statistical shape of the distribution.
We can see that the real distribution exhibits jagged edge. It is likely due to the converstion of the data to the float format.

The Figure~\ref{fig:pix_val} presents the overall pixel value distributions. Both real and generated images exhibit a similar range and align well at the lower and upper extremes. The generated distribution closely follows the real one but appears smoother, particularly at the lower end, where the sharp decline observed in real images is not fully reproduced. The steep decline at the low-intensity end of the distribution can be attributed to the large fraction of empty or near-zero pixels, while the sharp rise at the high-intensity end is primarily driven by pixel saturation in the images (see Figure~\ref{fig:ex_data}). Following this initial decrease, a secondary excess is observed, which is associated with diffuse emission and fainter pixels located at the outskirts of galaxies. The generated samples exhibit an enhanced contribution in this regime, indicating that the network produces a larger fraction of low-surface-brightness emission compared to the real data.

Figure~\ref{fig:mean} and Figure~\ref{fig:std} compare the mean and the standard deviation distributions. The overall agreement between real and generated images is strong, with both distributions displaying similar trends. However, the real images show a slight excess compared to the generated data, and the generated distribution contains high-value outliers that extend beyond the range observed in real images.

Overall, the comparison between real and generated radio galaxy images suggests that the DDPM effectively captures the general statistical properties of the data, producing images with similar distributions in pixel values, brightest pixel intensities, mean, and variance. The strong alignment between real and generated distributions indicates that the model successfully learns key features of radio galaxies, preserving their overall structure and characteristics. However, some discrepancies remain, particularly in the brightest pixel values and at extreme variance levels. The generated images exhibit smoother transitions and slightly reduced intensity, suggesting that while the DDPM models the underlying structure well, it may struggle to fully reproduce the most extreme features, such as the brightest compact sources or highly variable regions.

%#####################
\subsection{Source statistics}

\begin{figure}[h]
    \centering
     \includegraphics[width=0.9\linewidth,keepaspectratio]{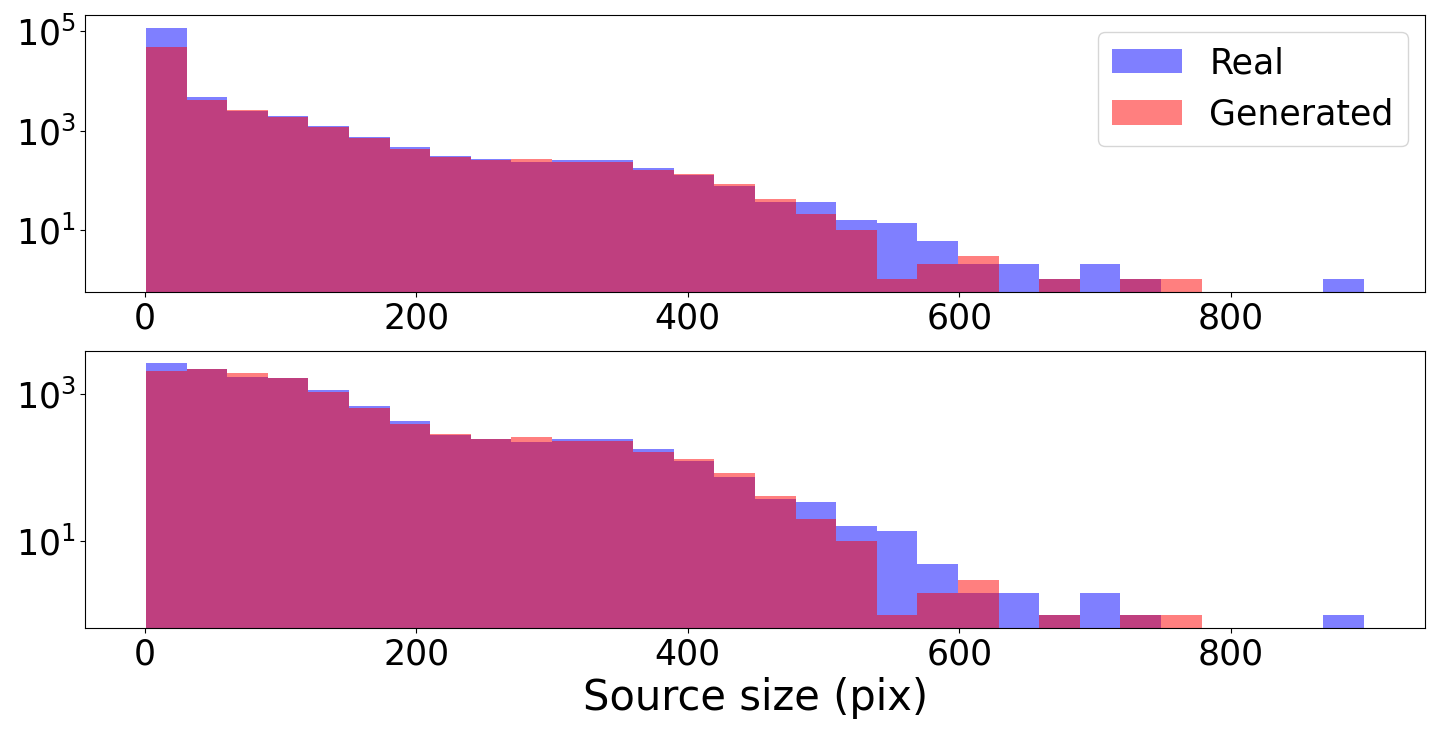}
    \caption{Statistical comparison of source sizes in the real data sample and those generated by our network. The top subplot shows the distributions of all sources, while the bottom subplot shows the distribution restricted to sources that contain at least one pixel with the maximum intensity in the image.}
    \label{fig:sources_size}
\end{figure}

\begin{figure}[h]
    \centering
     \includegraphics[width=0.9\linewidth,keepaspectratio]{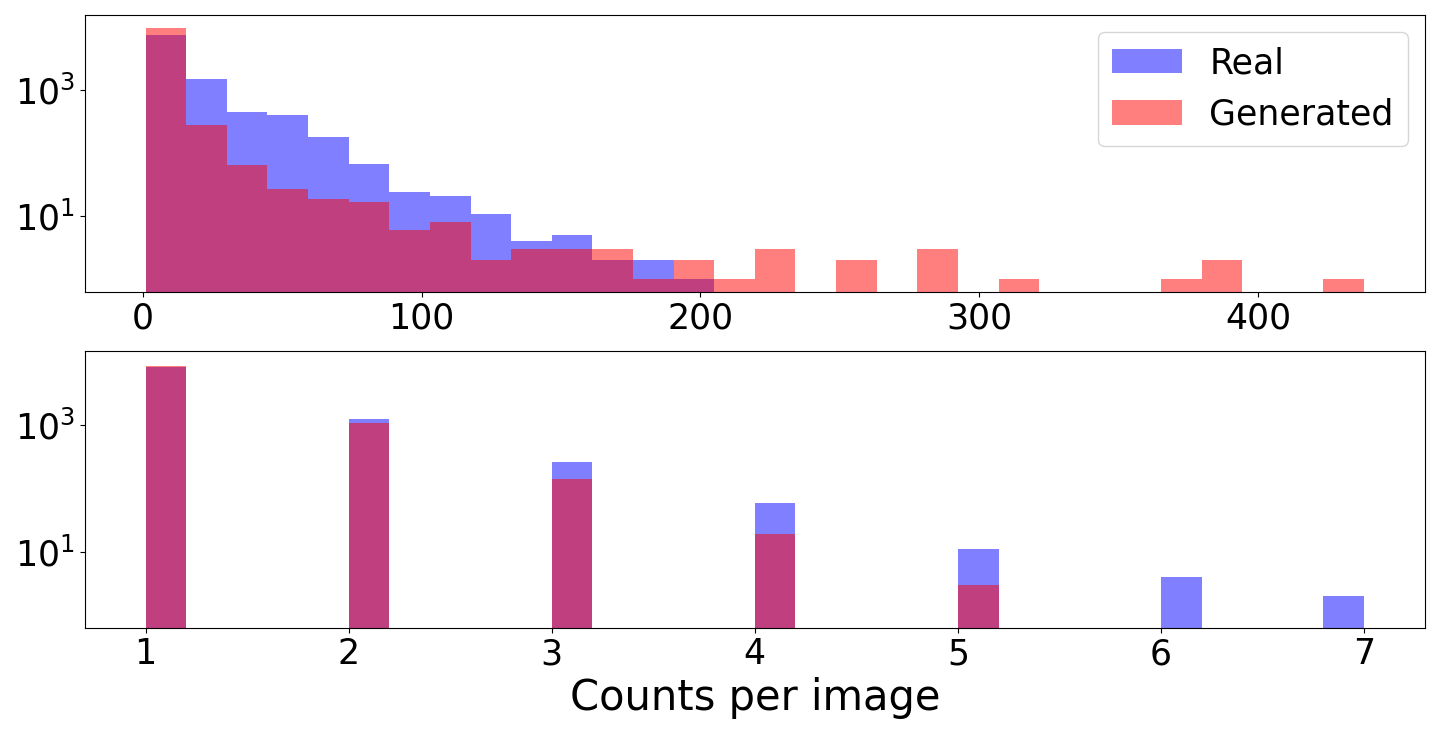}
    \caption{Statistical comparison of source counts in the real data sample and those generated by our network. The top subplot shows the distributions of all sources, while the bottom subplot shows the distribution restricted to sources that contain at least one pixel with the maximum intensity in the image.}
    \label{fig:sources_nbr}
\end{figure}

\begin{figure}[h]
    \centering
     \includegraphics[width=0.9\linewidth,keepaspectratio]{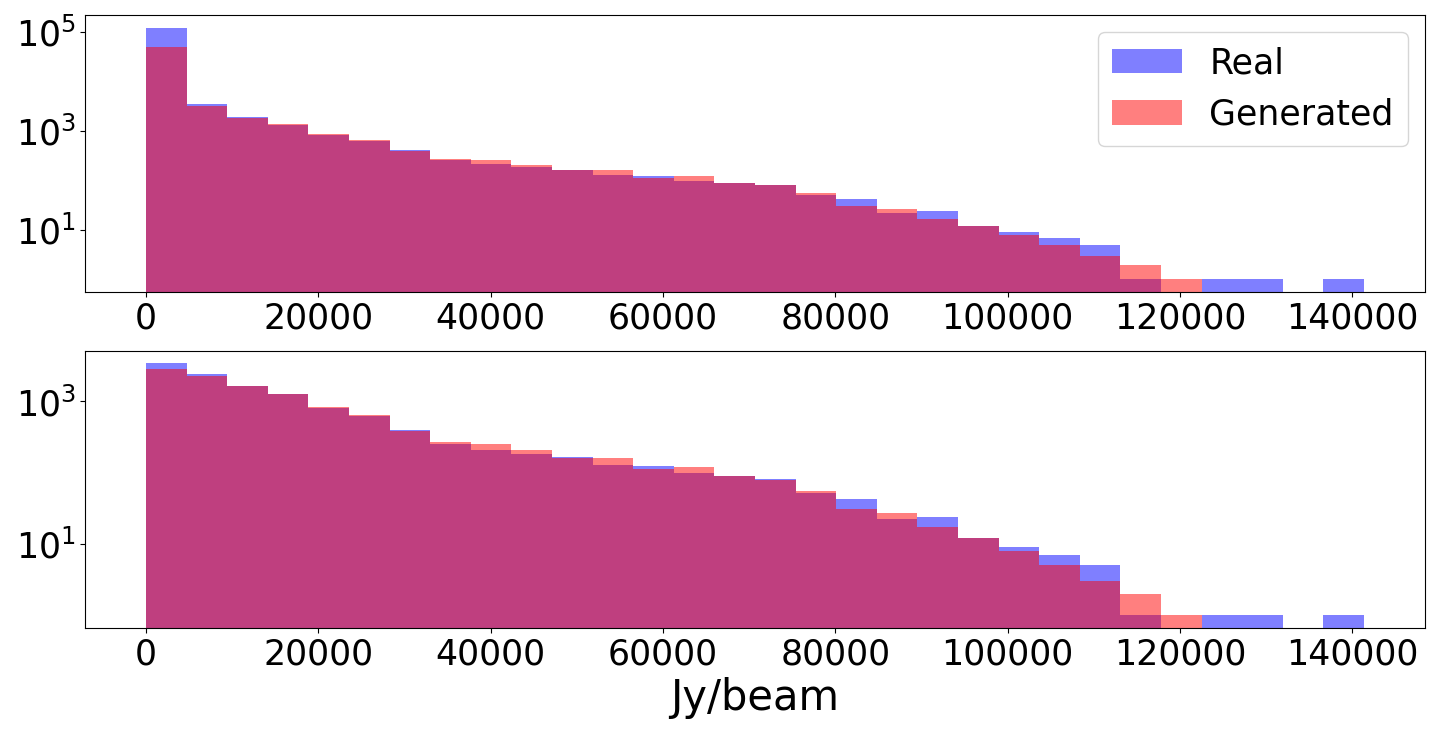}
    \caption{Statistical comparison of source intensities in the real data sample and those generated by our network. The top subplot shows the distributions of all sources, while the bottom subplot shows the distribution restricted to sources that contain at least one pixel with the maximum intensity in the image.}
    \label{fig:sources_intensity}
\end{figure}

The Figures~\ref{fig:sources_size},~\ref{fig:sources_nbr} and~\ref{fig:sources_intensity} consists of six subplots comparing statistical properties of real and generated sources. Each subplot presents a histogram with real data represented in blue and generated data in red. 
Respectively, we present the distributions of source sizes, the number of sources per image, and the total source intensity. The top subplots display the distributions for all detected sources in the images, while the bottom row focuses on sources that contain at least We use the same algorithm as the one used in section \ref{sect:prepro} to identify, count and determine the sources size in the generated images.

The model successfully captures the general statistical properties of real radio sources, producing source size and intensity distributions that broadly follow the expected trends. The generated sources exhibit similar peak positions and overall behaviors, demonstrating that the model is capable of learning key structural characteristics. However, notable discrepancies remain, particularly in the number of sources per image and the distribution of bright sources. The generated data show a more rapid drop-off in source sizes beyond 300 pixels and fail to reproduce the broader tail observed in real data, suggesting limitations in generating large and complex structures. Furthermore, the model struggles to generate small, faint sources, leading to an underestimation of the total number of sources per image. The excess of single-source images and occasional outputs dominated by noise indicate that the model does not fully capture the complexity of source clustering and background variations

%#####################
\subsection{Inpainting}

\begin{figure}[h]
    \centering
     \includegraphics[width=0.9\linewidth,keepaspectratio]{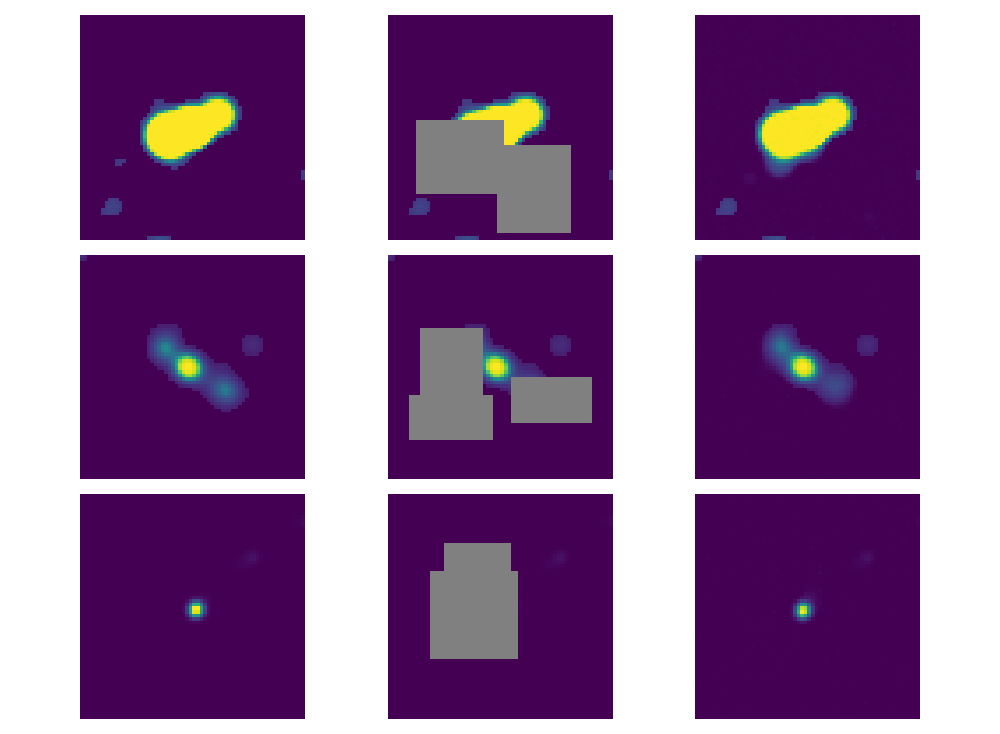}
    \caption{Examples of inpainting produced by our model. The left column shows the original images. The middle column shows the masked inputs given to the network, where gray regions mark the removed content. The right column shows the reconstructed images.}
    \label{fig:inp_ex}
\end{figure}

As discussed in section \ref{sec:guided}, our training method allows our model to generate and inpaint radio galaxy images. We inpainted 1000 images.We show some examples in Figure~\ref{fig:inp_ex}. 
Visual inspection indicates that our model inapinting results look realistic.

\begin{figure}[h]
    \centering
     \includegraphics[width=0.9\linewidth,keepaspectratio]{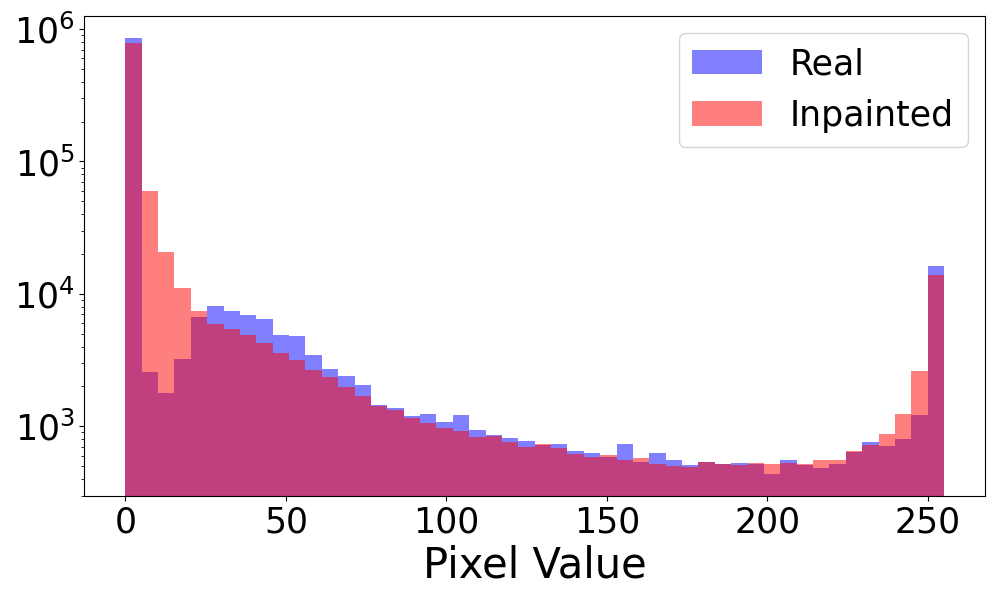}
    \caption{Statistical comparison of the pixel values distribution of the pixel values in the masked regions of the real data sample and the values of the inpainted pixels by our network.}
    \label{fig:inp_dist}
\end{figure}

\begin{figure}[h]
    \centering
     \includegraphics[width=0.9\linewidth,keepaspectratio]{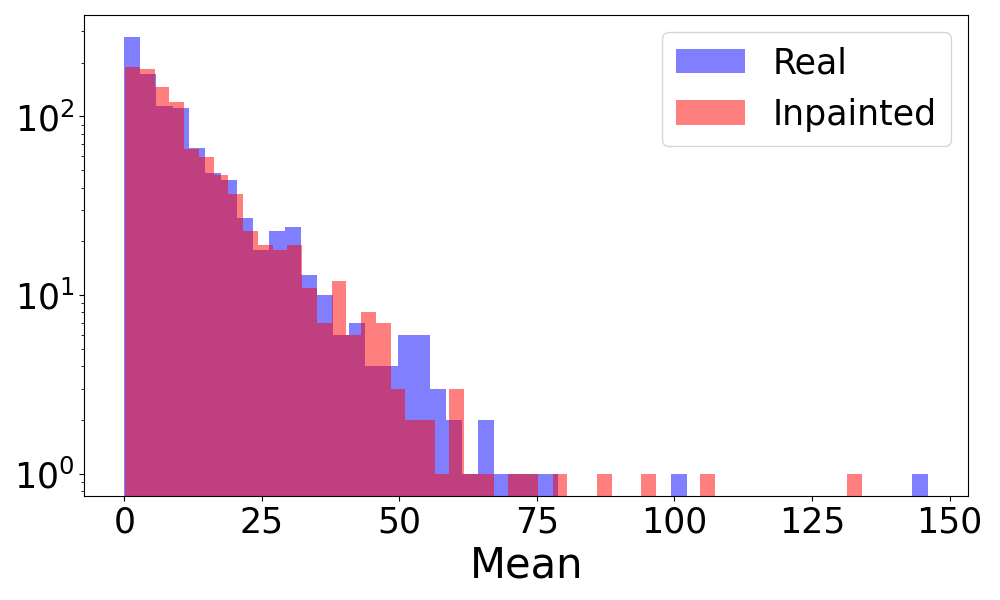}
    \caption{Statistical comparison of the pixel values distribution of the mean of masked region in each image.}
    \label{fig:inp_mean}
\end{figure}

\begin{figure}[h]
    \centering
     \includegraphics[width=0.9\linewidth,keepaspectratio]{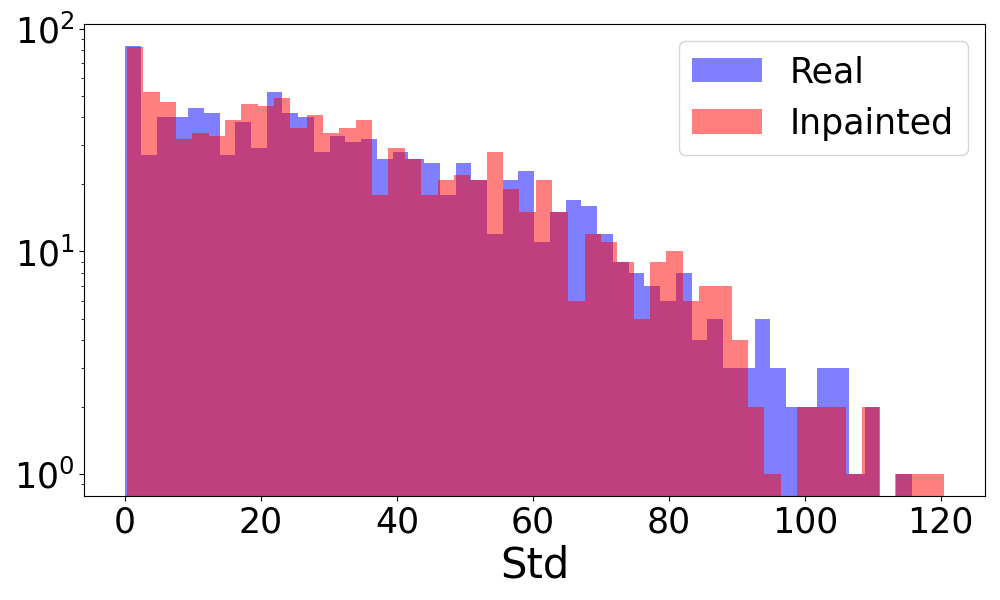}
    \caption{Statistical comparison of the pixel values distribution of the standard deviation of masked region in each image. The color code used is the same as Figure~\ref{fig:inp_dist}.}
    \label{fig:inp_std}
\end{figure}

To further quantify the quality of the inpainting capacity of our network, we looked at statistical distributions.
We first examine the distribution of inpainted pixel values relative to the original distribution, shown in Figure~\ref{fig:inp_dist}. The results are consistent with the trends seen in Figure~\ref{fig:pix_val}. The inpainted pixel distribution appears as a smoother version of the original one.

Figure~\ref{fig:inp_mean} shows the distribution of mean pixel intensities computed within the masked (and subsequently inpainted) regions. The blue histogram corresponds to the mean intensities measured from the same regions in the original images, while the red histogram shows those derived from the DDPM inpainted outputs. Both distributions exhibit a sharp decline, indicating that most masked regions contain low-intensity emission in both the real and reconstructed data.

Figure~\ref{fig:inp_std} presents the distribution of the standard deviation of pixel intensities within the masked regions. The blue histogram reflects the variability present in the original images, and the red histogram shows the variability recovered by the DDPM model. Both distributions span a wide range of values, demonstrating significant diversity in local structural complexity across the dataset. The substantial overlap between the two distributions indicates that the model generally preserves the intrinsic variability of the underlying emission.

Figure~\ref{fig:inp_ex} presents examples in which the DDPM successfully recovers features that are fully or partially masked. The reconstructed features include a large elliptical structure, the jet of a radio galaxy, and a point source.
A robust and detailed analysis of feature recovery would require either labeled datasets containing both the original and inpainted images, or a highly accurate method for classifying features in both datasets. At present, neither of these resources is available. Nevertheless, class-dependent structures should be indirectly reflected in the pixel-value distributions. Since no significant deviations are observed between the original and reconstructed distributions, there is no evidence that the class representation is being systematically distorted. This suggests that the DDPM is capable of recovering features correctly in a substantial fraction of cases.

In some cases, the masked regions fully obscure the original signal. When this occurs, the network receives no information from the underlying data and the output is no longer a true inpainting, but a generated reconstruction constrained only by the imposed mask size. This effect introduces a bias in the statistical distributions we analyze, although its magnitude cannot be quantified with the current approach.

%#######################################################
\section{Conclusions}\label{sect:conclu}

In this study, we successfully generated synthetic radio galaxy images that visually and statistically resemble real observations. To quantitatively assess the reliability of these images, we compared various statistical distributions, including pixel value distributions, the distribution of the brightest pixels, as well as the mean and variance of the generated and real datasets. The strong alignment between these distributions suggests that the model effectively captures key features of radio galaxies, preserving their overall structure and characteristics.

However, discrepancies remain, indicating that the DDPM struggles to fully reproduce certain complex behaviors observed in real radio galaxies. These differences highlight the need for further refinements, such as improved training strategies or more sophisticated conditioning techniques, to enhance the model’s ability to generate physically accurate representations of radio sources.

We also tested our DDPM performance with inpainted images. We saw that our model produce realistic looking results and that the statistical distributions of the inpainted regions have similar statistics.

In this study, we chose to use DDPMs due to their training stability and ability to generate high-quality images, the feasibility of preparing their high computational cost and slow image generation process as well as their capacity for inpainted images. Additionally, DDPMs serve as a foundation for training denoising diffusion restoration models \citep{2022kawar}, which extend their applicability to image restoration. This capability could be explored in future studies to enhance data quality and recovery in radio astronomy.

Moreover, most generative models have been developed and optimized for everyday life images, limiting their direct applicability to scientific domains. This is particularly relevant in astrophysics, where observations differ fundamentally from  everyday life images. Astrophysical images are often sparsely populated, with large regions containing little to no emission. Additionally, the physical properties being studied are encoded not only in the morphology of the observed structures but also in the precise pixel values, which carry essential scientific information. They also frequently exhibit diffuse emission, making it challenging to define clear object boundaries. Furthermore, the dynamic range of pixel values in astrophysical images can be significantly more extreme than in conventional datasets.

These differences highlight the need for adapting and refining generative models for astrophysical applications. Ensuring that DDPMs and other generative approaches can accurately capture both the structural and pixel-value distributions of astrophysical objects is crucial for their effective use in synthetic data generation for astrophysics. 

%######################################################

Future work could focus on optimizing the training process, incorporating domain-specific constraints, or leveraging hybrid generative approaches to improve both fidelity and diversity in the generated samples. 

Alternative approaches to DDPM can be considered to reduce computational cost at inference time. For instance, denoising diffusion implicit model \citep{DDIM} provides a deterministic sampling procedure that allows for a significant reduction in the number of denoising steps while maintaining comparable sample quality. Another promising direction is the use of latent diffusion model \citep{latent}, where the diffusion process is performed in a lower-dimensional latent space learned by an autoencoder, drastically reducing both memory usage and computational time. These approaches enable faster generation and facilitate scaling to higher-resolution data while preserving overall model performance.

We could also optimize the dataset construction. For example, we extract patches around the sources so that they are as centered as possible. This ensures that most of the emission is captured within each patch and simplifies the learning task. However, it may also introduce biases, such as an implicit preference for centrally located sources and reduced translational invariance. As a result, the model may learn overly simplified spatial priors and struggle to generalize to more complex or off-center configurations.

Additionally, exploring the use of DDPMs for tasks beyond image generation, such as denoising or super-resolution in radio astronomy, could further expand their applicability. As generative models continue to advance, refining their adaptation to astrophysical data will be crucial for maximizing their potential in large-scale surveys, such as those conducted with the SKA, where automated and reliable synthetic data generation could play a key role in data augmentation and analysis.

%#######################################################
\section{Data and code availability}\label{sect:data_avail}

The code for the data pre-processing and the DDPM are available  \href{https://github.com/hiboubbouter/Radio-Gal-DDPM}{here}.
The CIANNA framework is available \href{https://github.com/Deyht/CIANNA}{here}.
The FIRST dataset is available \href{https://zenodo.org/records/7351724}{here}.
The MGCLS cutouts will be available when Etsebeth (in prep) will be published.

%#######################################################
\section*{Acknowledgements}
RP acknowledges financial support from the SNSF under the Weave/Lead Agency project RadioClusters (214815). ET acknowledges financial support from the SNSF under the Starting Grant project Deep Waves (218396). This work was supported by EPFL through the use of the facilities of its Scientific IT and Application Support Center (SCITAS) VE acknowledges support from the South African Radio Astronomy Observatory and the National Research Foundation (NRF) towards this research. Opinions expressed and conclusions arrived at, are those of the authors and are not necessarily to be attributed to the NRF. This research made use of data from the MeerKAT telescope, a South African Radio Astronomy Observatory (SARAO) facility. SARAO is a facility of the National Research Foundation, an agency of the Department of Science and Innovation.

%#####################################################
%#######################################################%#######################################################
\FloatBarrier
\begin{appendix}
%#######################################################
\section{Residual Block}

\begin{table}[h]
\caption{\label{table:res_block}Residual Block}
\begin{center}
        \begin{tabulary}{\linewidth}{cccc}
                \toprule
                 Layer      & Filters     & Size  & Stride\\
                \midrule  
                 Group Norm      &   8         & -     &  \\
                 Conv            & $n$  & 3x3   & 1 \\
                 Group Norm      &   8      &  -    &  \\
                 Conv            & $n$/$\sqrt{2}$  & 3x3   & 1 \\
                 Conv            & $n$  & 1x1   & 1 \\
                 Merge           & -           & -     & - \\
                \bottomrule
        \end{tabulary}
\end{center}
\caption{Residual block used in the network describe in Sect.\ref{sect:archi}. Here, $n$ represent the number of filters chosen. For the Group Normalisation layer, the number of filters represents the group size parameter.}
\end{table}

\end{appendix}

\bibliographystyle{elsarticle-harv} 
\bibliography{bibli}



%% For example:

%% \bibitem[Aladro et al.(2015)]{Aladro15} Aladro, R., Martín, S., Riquelme, D., et al. 2015, \aas, 579, A101

%%\end{thebibliography}

\end{document}